# $T_c$-oscillations of a superconducting heterostructure OC/V/Fe by oscillating couple in a fragment OC=Fe/Cr/Fe.

Yu.V.Goryunov
*E.K. Zavoisky Physico-Technical Institute, Kazan, Russia*

**Abstract.**
Superconducting transition temperature ($T_c$) and FMR were measured in heterostructure OC/V/Fe, where OC=Fe/Cr/Fe. Giant oscillations of $T_c$ depending on a Cr-thickness were detected. Probable reasons of $T_c$ oscillations are discussed - periodic cancellation of paire-breacking exchange fields in superconducting V-layer or in fragment Fe/Cr/Fe.

Perhaps the most of significant achievements in investigation of properties of a multilayer hetero-structures a superconductor - ferromagnet (S/F) from the moment of a jolt given by the Radovic's theory [1] were the experimental evidence of existence of the π-phase superconductivity [2], prediction [3] and detection [4] of the effect of re-entrant superconductivity in S/F system [5]. At the present stage of experiment developments in the given area, it become actual check out predicted by the theory [6-8] the cancellation effects of exchange fields for Cooper pairs in layers forming supercon-ducting heterostructures. The study of these effects may bring to creation of a new device of superconducting electronics - spin valve.

In work [9] by means of polarization electron microscopy the veering of a magnetization of upper Fe-layer was visually shown depending on its location on a wedge of chrome layer, sputtered on uniformly magneti-zed bulk Fe single crystal. In beat, the given scheme is used in our attempt to demonstrate the influence of mutual directions of magnetizations of F-layers covering S-layer, only that as bulk Fe single crystal the same thin Fe epilayer is used.

Parameters of the prepared samples are shown on table 1. Samples were prepared by rf sputtering. The basic vacuum made $1.5\ 10^{-7}$ mbar. Producing pressure of a pure argon (99.999 %) during deposition is supported at a level $5\ 10^{-3}$ mbar. As targets the pure vanadium (99.99%), iron (99.99%) and electrolytic chrome (99.99% ) were utillized. As substrates for main samples series the monocrystal plates of MgO(001) with a roughness ~2 Å were utillized. Before sputtering the substrates and targets were exposed to the treatment in the radio-frequency discharge. At deposition the temperature of substrates was supported 300 °C, the rate of vanadium deposition was 0.6 Å/s at power 500W, iron and chrome ~ 0.1 Å/s at power 300W, recognized in a works [10,11] to be optimum for obtaining the most sleek with perfect crystalline structure of layers and, accordingly, sharp interfaces.

For preparation of wedge layers the targets of iron or chrome with a diameter 100 mm were used. Distance up to substrates made 77 mm. The 10-11 substrates of width 3 mm were placed symmetrically concerning a vertical axis of a target in a window with the size 8x33 mm². The long axis of a window was perpendicular to direct edge of the screen located at height of 40 mm above a target, and in 14 mm from a vertical axis of a target. The thickness distribution was measured on the specially prepared wedge with max thickness 200 Å Fe, by means of a optical density measurement at λ=660 nm as in [12].

The thickness distribution in the location of substrates relatively a central point was well featured by an expression: $D(x)/D_{max}= 0.368-0.0311\ x +7.33\ 10^{-4}\ x^2$. The precision of a thickness definition in a given place of a wedge is determined by reproducibility at each instal-lation of a substrate holder concerning an edge of the screen (0.5 mm) and is about 2 % from maximal thickness of a wedge.

For preparation of basic layers with a homogeneous thickness a target with diameter 100mm (iron) or 150 mm (vanadium) was utillized. The thickness distribution in the location of substrates relatively a central point was close to distribution of thickness of a Lambert radiant and was well featured by an expressions:
$D(x)/D_{max}=1-1.11\ 10^{-4}\ x^2$ (100 mm target) and
$D(x)/D_{max}=1-7.17\ 10^{-5}\ x^2$ (target with diameter 150 mm).
Problem a thickness distribution for sputter technology is shown in detail in [12].

Since an impurity content defines a coherence length, which defines in its turn critical and requiring thickness of S-layer, all samples were prepared with the same tuning up of sputtering equipment and usage of the same targets. So the main seriess S976 and S982 have identical quality of S-layers.

**The superconducting transition temperature** of samples without the Cr-layer was measured resistively in a four-terminal configuration. The influence of a V-layer thickness on $T_c$ of a Fe/V/Fe heterostructure was investi-gated on samples of series S956 (tab.1). These measure-ments have shown qualitatively earlier known [1,13-15] $T_c(d_V)$ dependence and have allowed precisely to deter-mine a working interval of thicknesses of V- layer for the seriess S976 and S982. The reference characteristic values of this dependence were obtained as the follows: $d_V^{crit}$ ~290 Å, $T_c(d_V\ \sim 5\ d_V^{crit}) = 4.35$ K. The similar

**The table 1. Parameters of the prepared sample series.**

| The series number | Substrate, alternation of layers, protection layer | S - layer | | F - layer | | AF-layer | | Residual Resistiv. Ratio, **RRR** |
|---|---|---|---|---|---|---|---|---|
| | | Thickness, Å | Dep. rate, Å/s | Thickness, Å | Dep. rate, Å/s | Thickness, Å | Dep. rate, Å/s | |
| **S 956** | glass / Fe /V/ Fe 20 Å Pd | 250 – 1300, Vanad. | 0.35 | 15, 15 | 0.16 | - | - | 4 |
| **S 976** | MgO (001) / Fe/V/Fe 20 Å Pd | 335, Vanad. | 0.6 | a wedge 3-34 | max 0.1 | - | - | 12 |
| **S 979** | $Al_2O_3$ (1120) / Fe/Cr/Nb/Cr/Fe 30 Å Nb | 380, Niob. | 0.3 | 30 | 0.1 | a wedge, 3-30 | max 0.15 | 3 |
| **S 982** | MgO (001) / Fe/Cr/Fe/V/Fe 60 Å V | 340, Vanad. | 0.6 | 8, 8, 20 | 0.1 | a wedge, 0.8-15 | max 0.15 | 12 |

dependence for system V/Fe, obtained in other conditions is given in [5]. From dependence $T_c(d_V)$ for a series S956 follows, that $T_c$ deviation in a series S976 and S982 due to inhomogeneity thickness of V-layer for peripheral samples does not exceed 0.13 K.

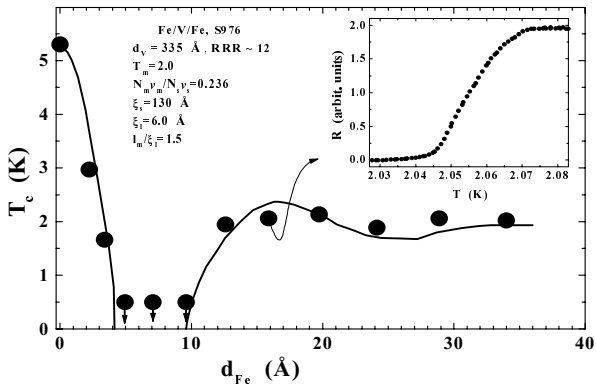

FIG. 1. $T_c$ vs $d_{Fe}$ of a epitaxial heterostructure MgO(001)/Fe/V/Fe (series S976 from work [4] or series 6 from work [5]). A line is the L.Tagirov's calculation at parameters of the theory shown in figure. Coherence length for vanadium evaluated from data on a resistance (RRR ~ 12) is 130 Å. Example of superconducting transition curves measured by electrical resistivity is shown on the insert

The $T_c$ dependence on a Fe-layer thickness was investigated on samples of series S976 with the wedge iron layers. The typical features of $T_c(d_{Fe})$ dependence were: primary slump of $T_c$, lack of superconductivity for three samples with thickness in an interval 4-11 Å, and the increasing of $T_c$ to a level $T_c \sim 2.0$ K for the thicker Fe-layers. The given dependence, which was published in [4] and [5] (series 6) together its course, calculated by L.Tagirov, is shown once again in a fig.1. Thus effect of re-entrant superconductivity was detected and the best value of the iron layer thickness (6-9 Å) was determined, which utilised in the present work.

The $T_c$ of samples with a Cr-layer was determined from magnetic susceptibility measurements on an alternating current with 6 MHz frequency. In these measurements sample of the size 3x8 $mm^2$ put on the flat measuring coil with diameter 6 mm. As the thickness of a S-layer and the size of all samples were the same, that this signal with precision of ~ 10 % was proportional to the area of a superconducting sample part. Errors of the temperature measurements are estimated: systematic deviation is 0.05 K, random error is 0.001 K.

The influence of a separating interlayer of chrome between a F–layer and S-layer of a heterostructure F/S/F was investigated on example of S979 series of Fe/Nb-system. It was established, that the Cr-layer by thickness more than 10 Å completely isolates a F-layer from a S-layer. The results of these measurements for a series S979 ($Al_2O_3$(1120) /Fe/Cr/Nb/Cr/Fe) are represented in a fig.2. The magnetic and superconducting properties of Fe/Nb layer system was investigated in work [16].

The ac measurements of S982 samples, containing a fragment Fe/Cr/Fe with a wedge Cr-layer were executed along all wedge of 30 mm with a step of 3 mm. Superconducting transitions for each of 10 samples of S982 series are represented in figure 3. It is seen in figure 3, that the transitions have various character: from sufficiently sharp transitions (samples - 1,2,3,8) up to broad (5) and double-transitions (6,10). The fact of observation for some samples of narrow transitions (see insert of fig.1 and fig.3) allows to conclude, that the broadening and diffusion of transitions are caused by $T_c$ distribution along the sample. Apparently, the given $T_c$ distribution is determined by monotonous distribution of thickness in a Cr-layer and the interval of values of a diamagnetic susceptibility, proportional to superconducting areas, is bound to an interval of Cr-thickness of each sample. In figure 4 the same superconducting transitions with ac- susceptibility expressed in Cr-thickness units are represented. These transitions locate on the relevant intervals of a Cr-layer thickness, taking into account, that the dependence of $T_c(d_{Cr})$ on Cr-thickness should be almost monotonous in an interval of effective thickness less than Cr-monolayer according to reasons of spatial crystalline symmetry. As the etalon the average value on 1-3 samples is taken for recalculation of a ac- susceptibility arbitrary units in unities of Cr-thickness. As each transition should occupy all region of Cr-thickness relevant to a sample, a missing part of a diamagnetic susceptibility for each sample is presented by a virtual transition with $T_c$ lower than a temperature limit of a cryostat equal to 1.5 K. Thus, the summary curve shown by full circles represents the $T_c(d_{Cr})$ dependence on Cr-thickness. This representation is distorted for samples with the thickest Cr-layer, since requirement of smallness of a Cr-thickness interval is violated.

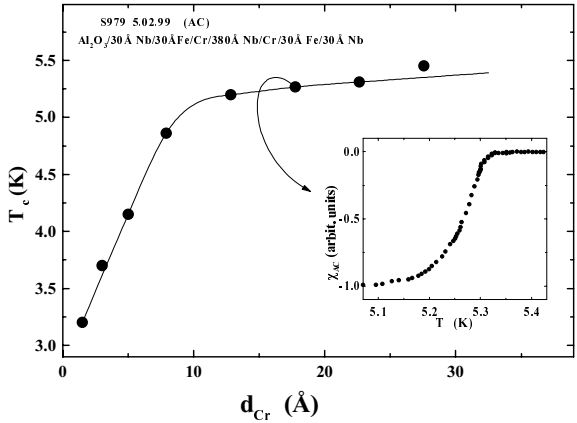

FIG. 2. $T_c$ vs $d_{Cr}$ of a epitaxial heterostructure $Al_2O_3/Fe/Cr/Nb/Cr/Fe$ (series S979). Line is lead for "eye". Example of superconducting transition is shown on the insert.

**FMR measurements** were carried out for selective estimation of a magnetic state of individual Fe-layers and a fragment Fe/Cr/Fe, included in a heterostructure of series S976 and S982. The measurements were conducted at a frequency 9.4 GHz in the $TE_{102}$ cavity at room temperature. For determination of mutual magnetic axis directions of Fe-layers an angular dependences of resonance field were measured. The measuring technique and common FMR features in a single Fe-layer on MgO-substrate are stated in [17]. FMR measurements, carried out on a series S976, have confirmed behaviour of spectrums and angular dependences of single Fe-layer in MgO(001)/V/Fe systems, observed in [17], and have allowed to make precise identification of signals FMR on layers of the series S982. We observed in given series, that a FMR line intensivity of fragment Fe/Cr/Fe for Cr - thickness near two Cr monolayers is ten times decreased in comparison when thickness of 1 and 3 Cr-monolayers, i.e fragment Fe/Cr/Fe ceases to be ferromagnetic. For greater Cr-thickness the decreasing of a FMR-line inten-sivity was observed also. We connect this decreasing with strong anisotropy of thrilayer Fe/Cr/Fe with thick Cr-layer, making the FMR -signal unobservable on frequency 9.4 GHz. [18]. Such behaviour is well conformed to dates [8,18,19] for samples grown at the substrate temperature 300°C and calculations [20], showing that the period of oscillating couple in Fe/Cr/Fe thrilayer is near two Cr-monolayers, magnetizations of layers are homogeneous and either parallel or antiparallel. Basing on this one can conclude, that superconducting area on all surface of S982 corresponds to area with antiparallel mutual arrangement of magnetizations of Fe-layers in a Fe/Cr/Fe fragment. It is important to note, that detected by us the change of view of a resonance field angular dependence indicates on switching of a light magnetization axis from a direction Fe [001] on a direction Fe [011] for a small resonating part of a sample with Cr-thickness more 4 monolayer. Noncollinearity [21] of mutual position of magnetizations of F-layers can yield the contribution to a shape of $T_c(d_{Cr})$ dependence.

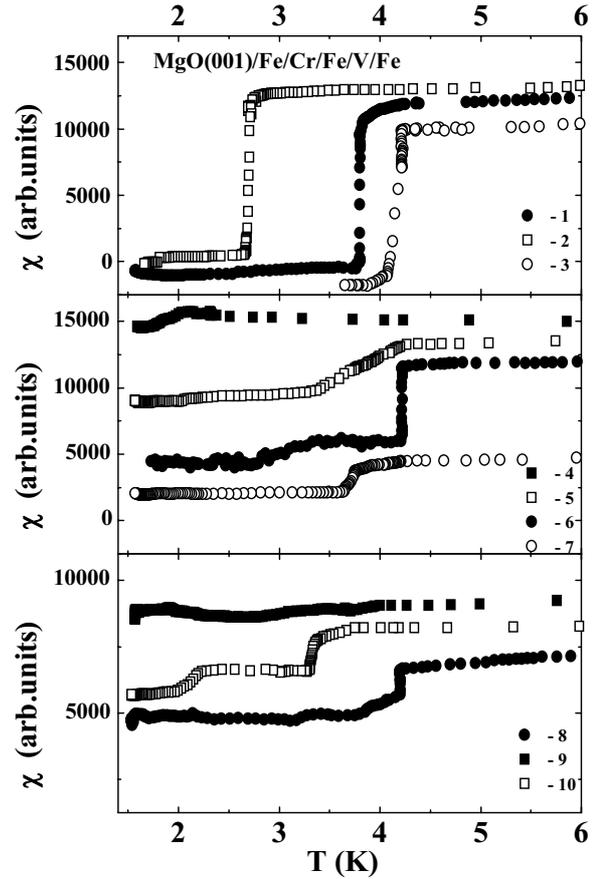

FIG. 3. Superconducting transition curves for all samples of heterostructures OC/V/Fe (OC=Fe/Cr/Fe) (series S982) measured by ac susceptibility.

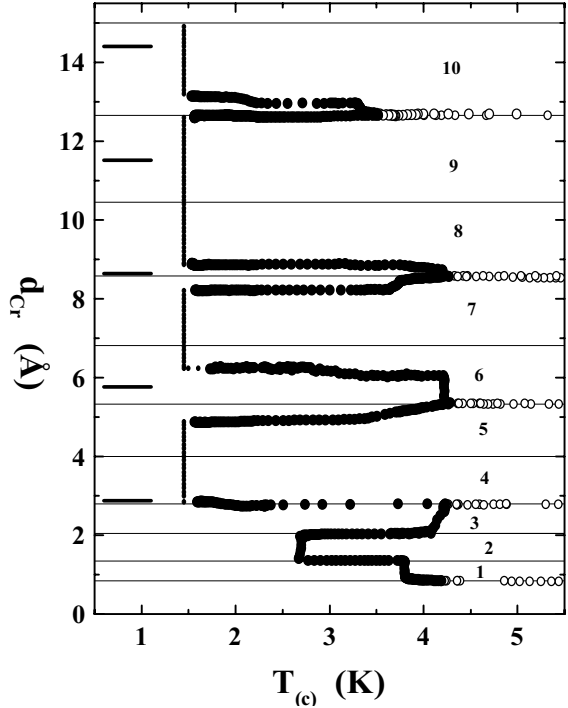

Fig. 4. Superconducting transition curves for all samples of heterostructures OC/V/Fe (OC=Fe/Cr/Fe) (series S982) from fig.3 in a Cr-thickness units. Full circles relevant to superconducting phase. Small full circles show invisible part of superconducting transition curves below the temperature limit of cryostat. The fine hori-zontal separate intervals of the Cr-thickness concerning to various samples. The fat horizontal stretchs show values of Cr-thickness relevant to antiferromagnetic orientation of magnetizations in fragment Fe/Cr/Fe.

In our opinion the fact of detection in explored samples of oscillations $T_c(d_{Cr})$ depending on a Cr-thickness does not call doubts. Possible reasons of observed $T_c$ - oscillations may be the following:

**A.** The display of effect similar to cryptoferromagnetizm [22,23,15]: the coexisting of ferromagnetism and superconductivity on different scales of lengths, i.e. when on the size of Cooper pair a great quantity of shallow ferromagnetic domains with opposite directions of magnetizations is stacked. If the magnetizations of Fe-layers in a fragment Fe/Cr/Fe are opposite, this fragment of a heterostructure does not render any influence on superconductivity of a vanadium layer, as if in the magnetic relation would not exist, and all system of layers can be represented as bilayer of a superconductor - ferromagnet. In case of bilayer the theory and experiment [13] display, that its $T_c$ is equal $T_c$ of thrilayer F/S/F with double thickness of a S-layer, i.e. in this case according to $T_c(d_V)$ dependence, represented by series S956 (or series 1 in [5]), the essential increase of $T_c$ should be observed. It is possible, that such behaviour of $T_c$ is bound to features of shaping in chrome of a spin density wave and change of conditions of reflection of Cooper pairs on the interface. But data of works [24], in which the influence of chrome on super-conductivity of niobium and vanadium was studied, point that chrome in layers less than 30 Å practically does not influence their superconductivity. Data in fig.2 about a low transmittivity for Cooper pairs of a Cr-layer thicker 10 Å against this version. It increments probability of the following version.

**B.** Another possible explanation of observed $T_c$ oscillations can be a cancellation effect of an opposite directional exchange fields on a vanadium layer or the effect of a spin valve [7,26]. If to estimate a coherence length of a V-layer from resistance data, we have $d_V/\xi_S = 2.6$. It specifies on the inapplicability of the Tagirov's theory in this case. But as the behaviour $T_c$ of a series S976 and S982 will match qualitatively to behaviour featured in [7], we suppose, that this description requires improvement in a part of application of the concepts of spatial, time and phase coherence lengths. (For example, it was shown in Mercereau experiments [25], that the phase coherence is maintained on distances more than 1 m). In the theory of a proximity effect, using an optical analogy with a Fabry-Perot interferometer, the concepts of time and spatial lengths of a coherence are identical. In case of pure superconductors it is also true. But in case of dirty superconductors, in our opinion, there occur distinctions. In fact a time coherence of superconducting wave function is influenced primarily by pair-breaking spin-up scattering processes. Therefore at calculation of length of a coherence it is necessary to take into account a free spin diffusion length of electrons. This value can be compared to thickness of a S-layer, i.e. already in a normal state there are backgrounds of cancellation of pair-breaking exchange fields, created in a S-layer by a F-layers.

Thus, the observation of a cancellation effect exchange fields for Cooper pairs in the given heterostructure confirms an opportunity of manufacture of a spin valve on a basis, at least of one of combinations of two alternatives: The first alternative: either **1a**- interior (in relation to a S-layer) cancellation of exchange fields or **1b**- exterior [23] cancellation of exchange fields. The second alternative: either **2a**- the thickness of a S-layer is more than critical thickness ($d_{crit}$) or **2b**- the thickness of a S-layer is less than $d_{crit}$. Combination (**1a, 2a**) is the version calculated by L.Tagirov [7].

The author would like to thank the Director of a solid state physics of the Ruhr-Bochum university Professor H.Zabel for the submitted opportunity of sample preparation with its growth facilities, colleagues for the valuable notes and the Russian Fund for Basic Research for support of this work (Projects No.03-02-96191a and No.03-02-16382)

Corresponding author: Yu.V.Goryunov
E.K. Zavoisky Physic-Technical Institute.
Sibirskii tract,10/7, 420029, Kazan, Russia
E.mail: gorjunov@kfti.knc.ru